\documentclass[pdflatex,sn-mathphys-num]{sn-jnl}

\usepackage{graphicx}%
\usepackage{multirow}%
\usepackage{amsmath,amssymb,amsfonts}%
\usepackage{amsthm}%
\usepackage{mathrsfs}%
\usepackage[title]{appendix}%
\usepackage{xcolor}%
\usepackage{textcomp}%
\usepackage{manyfoot}%
\usepackage{booktabs}%
\usepackage{algorithm}%
\usepackage{algorithmicx}%
\usepackage{algpseudocode}%
\usepackage{listings}%

\theoremstyle{thmstyleone}%
\theoremstyle{thmstyletwo}%

\theoremstyle{thmstylethree}%

\begin{document}

\title[Article Title]{DDNet: A Unified Physics-Informed Deep Learning Framework for Semiconductor Device Modeling}


\author[1]{\fnm{Roberto} \sur{Riganti}}

\author[2]{\fnm{Matteo G. C.} \sur{Alasio}}

\author[3,4]{\fnm{Enrico} \sur{Bellotti}}

\author*[1,3,4]{\fnm{Luca} \sur{Dal Negro}}\email{dalnegro@bu.edu}

\affil[1]{\orgdiv{Department of Physics}, \orgname{Boston University}, \orgaddress{\street{590 Commonwealth Avenue}, \city{Boston}, \postcode{02446}, \country{USA}}}

\affil[2]{\orgdiv{Dipartimento di Elettronica e Telecomunicazioni}, \orgname{Politecnico di Torino}, \orgaddress{\street{Corso Duca Degli Abruzzi 24}, \city{Torino}, \postcode{10129}, \state{TO}, \country{Italy}}}

\affil[3]{\orgdiv{Department of Electrical and Computer Engineering and Photonics Center}, \orgname{Boston University}, \orgaddress{\street{8 Saint Mary’s Street}, \city{Boston}, \postcode{02215}, \state{MA}, \country{USA}}}

\affil[4]{\orgdiv{Division of Material Science and Engineering}, \orgname{Boston University}, \orgaddress{\street{15 Saint Mary’s Street}, \city{Brookline}, \postcode{02446}, \state{MA}, \country{USA}}}


\abstract{
The accurate modeling of semiconductor devices plays a critical role in the development of new technology nodes and next-generation devices. Semiconductor device designers largely rely on advanced simulation software to solve drift-diffusion equations, a coupled system of non-linear partial differential equations that describe carrier transport in semiconductor devices. Although these tools perform well for forward modeling, they are not suitable for addressing inverse problems, for example, determining doping profiles, material properties, and geometric parameters given desired device performance. Meanwhile, physics-informed neural networks (PINNs) have grown in popularity in recent years due to their ability to efficiently solve inverse problems at minimal computational cost compared to forward problems. In this study, we introduce the Drift-Diffusion Network (DDNet), a unified physics-informed deep learning solver for forward and inverse mesh-free solutions of the drift-diffusion equations. Using prototypical device configurations in one and two spatial dimensions, we show that DDNet achieves low error compared to traditional simulation software, while additionally solving user-defined inverse problems with minimal computational overhead. We expect that DDNet will benefit semiconductor device modeling by enabling the inverse design paradigm of semiconductor devices from any user-defined objectives, such as electric fields or charge distributions, thus facilitating the discovery of novel structures in comprehensive parameter sets and in a fully automated way.}





\maketitle

\section{Introduction}
Semiconductor devices are fundamental components of modern microelectronics, enabling the operation of integrated circuits, optoelectronic systems, and power management technologies~\cite{sze_physics_2021}. Their behavior, governed by charge carrier transport at the microscale~\cite{2016Ferry}, directly impacts the performance, energy efficiency, and scalability of computing, communication, and sensing technologies~\cite{xie_designing_2024, zhao_review_2024, thennarasu_comprehensive_2025}.

As device architectures continue to scale and diversify, precise modeling of semiconductor physics remains essential for both scientific understanding and technological development. Among the various approaches, the drift-diffusion model~\cite{1985Markowich, 2012Markowich, 2008Vasileska}, consisting of a set of coupled parabolic partial differential equations (PDEs), forms the backbone of most modern commercial simulation tools. These transport equations, typically grouped into Poisson and continuity equations, describe the transport of charge carriers under external excitation. Due to the intrinsic non-linearity of the equations, analytical solutions are only possible in limited cases, and numerical approaches are required~\cite{1983BankRose, 1985Markowich, 2012Markowich}.

Commercial software packages such as Synopsys TCAD Sentaurus~\cite{Sentaurus_Device_W-2024.09-SP1}, Silvaco Atlas~\cite{silvaco_sftw}, and COMSOL Multiphysics~\cite{comsol_sftw} implement advanced numerical solvers based on finite element or finite volume methods. These tools provide high accuracy in 1D, 2D and 3D simulations, support a wide range of physical models and boundary conditions, and are widely used for device design, optimization, and analysis in both academia and industry.

Despite their maturity, traditional solvers face several limitations. Their convergence often depends on the quality of the mesh, numerical approximations of differential operators, initial guesses, and extensive parameter tuning. Furthermore, solving coupled non-linear PDEs typically requires Newton-type second-order methods, which are highly sensitive to initialization choices~\cite{1983BankRose}. Lastly, exploring high-dimensional parameter spaces, solving inverse problems, or integrating sparse experimental data often exceeds the capabilities of conventional solvers. These challenges motivate the development of an alternative paradigm that takes advantage of the flexibility and power of artificial neural networks (ANNs) and provides physical consistency, scalability, parameter training capabilities, and the fast solution of user-defined inverse design problems~\cite{raissi_physics-informed_2019}. 

In this context, recent advances in scientific machine learning have introduced physics-informed neural networks (PINNs) as a mesh-free alternative to solve forward and inverse problems governed by PDEs~\cite{raissi_physics-informed_2019, chen_physics-informed_2020,eivazi_physics-informed_2022, chen_physics-informed_2022, riganti_auxiliary_2023}. Unlike conventional data-driven neural networks, PINNs enforce the validity of physical laws by embedding the governing equations directly into the loss function. These models use feedforward neural network architectures as trainable surrogates of PDE solutions and leverage automatic differentiation (AD) to compute PDE residuals at randomly distributed collocation points in the solution domain and on its boundary~\cite{haykin_neural_2009, goodfellow_deep_2016,lu_deepxde_2021}.

In the case of forward simulations, PINNs eliminate the need for labeled data, relying solely on physics-based constraints. This makes them well-suited for problems where data are scarce or unavailable. Furthermore, inverse problems~\cite{tarantola_inverse_2005, chen_computational_2018} can be handled by simply introducing an additional loss function, without significant computational overhead~\cite{riganti_multiscale_2025, chen_physics-informed_2020, chen_physics-informed_2022}. These characteristics make PINNs a flexible and powerful simulation framework for solving a wide range of forward and inverse differential and integro-differential problems~\cite{raissi_physics-informed_2019, chen_physics-informed_2020,eivazi_physics-informed_2022, chen_physics-informed_2022, riganti_auxiliary_2023, mao_physics-informed_2020, raissi_hidden_2020, yu_gradient-enhanced_2022, xu_understanding_2025}.

With respect to semiconductor device modeling, limited progress has been achieved so far using PINN methods~\cite{li_overview_2024}. Cao et al. investigated PINNs for Poisson-like equations, without, however, extending their results to relevant electrical engineering problems~\cite{cao_physics-informed_2023}. Cai et al. proposed a TCAD forward simulation PINN-based architecture to solve a one-dimensional $p$-$n$ junction simulation. However, they did not discuss the generality of their method in higher dimensions or for inverse problems~\cite{cai_multi-order_2024}. Kim and Shin used a PINN-DeepONet hybrid solver to simulate the electrical characteristics of a one-dimensional device based on the TCAD training data~\cite{kim_novel_2023}. Lastly, Liu et al. developed a one-dimensional Poisson-Boltzmann solver, which addressed parameter retrieval problems~\cite{liu_asymptotic-preserving_2025}. As a result, there remains a critical need for a forward and inverse (unified) PINN framework that can address relevant devices and coupled simulations with inverse design capabilities. Such a framework should match the precision and reliability of state-of-the-art industrial solvers while demonstrating greater flexibility and scalability in integrating different boundary conditions, generation-recombination rates, and inversely retrieving relevant device parameters. 
Moreover, an innovative physics-informed deep learning framework should also demonstrate a clear computational advantage in solving relevant non-linear inverse problems beyond the reach of current numerical methods~\cite{noauthor_machine_2025}. As recently pointed out, PINNs should not be employed to solve relatively simple forward problems against a weak baseline without demonstrating a clear computational advantage over traditional solvers~\cite{mcgreivy_weak_2024}. Specifically, we argue that the crucial advantage of PINNs compared to traditional solvers stems from their inherent ability to successfully tackle inverse design problems where physics-constrained solutions, representing targeted functionalities or device configurations, are efficiently recovered even in the case of stiff and ill-posed problems~\cite{lu_physics-informed_2021, riganti_multiscale_2025}. Furthermore, there have been increasing efforts to extend the applications of physics-based surrogate models to benchmark performance applications, such as in the automotive aerodynamics field~\cite{tangsali_benchmarking_2025}, stressing the importance of physics-informed deep learning for industrial applications.

In addition, fields such as fluid dynamics and electromagnetics have experienced substantial advances in physics-informed surrogate modeling over the past five years, despite the availability of established forward Navier-Stokes or electrodynamics numerical solvers~\cite{comsol_sftw,chen_physics-informed_2020,eivazi_physics-informed_2022,chen_physics-informed_2022,riganti_multiscale_2025, mao_physics-informed_2020, raissi_hidden_2020}. In contrast, the comparatively limited progress of PINNs in semiconductor device modeling originates from the intrinsic challenges associated with the highly non-linear nature of the drift–diffusion equations, whose solutions present large gradients and span extremely large dynamic ranges. In our view, these considerations underscore the need for fundamentally new network architectures and training methodologies. 

In this work, we propose the \textit{Drift-Diffusion Network (DDNet)}, a unified physics-informed neural network architecture for the solution of both the forward and inverse design problems of semiconductor device modeling. In particular, we demonstrate that DDNet reproduces forward solutions obtained with commercial solvers and solves inverse design problems at a minimal additional cost. Through multiple examples involving one-dimensional (1D) and two-dimensional (2D) device geometries, we establish the accuracy of DDNet solutions by comparing against Synopsis TCAD Sentaurus simulations using the relative $L_1$ error metric defined in the Supplementary Information. Our results achieve accuracy comparable to conventional solvers in forward 1D and 2D device simulations, requiring only minimal network modifications. Finally, we showcase the parametric learning capabilities of DDNet and demonstrate its unique advantages for inverse device design over traditional numerical methods by efficiently recovering parameter-dependent functional responses and optimal doping profiles from target electric field distributions, with applications ranging from optoelectronics to sensing technologies.

While DDNet meets the essential benchmarks for comparison with current state-of-the-art modeling tools, it is not intended as a \textit{tout court} replacement of established numerical methods. Instead, our comparisons should be regarded as seminal steps toward the development of PINN-driven semiconductor device modeling techniques enabling device discovery at minimal computational costs.
Ultimately, our results unleash the potential of DDNet as a flexible and generalizable framework for microelectronics research, inverse design, and device discovery.

\section{Results}
\subsection{The DDNet framework for semiconductor modeling}
\subsubsection{The Poisson Drift-Diffusion Equations}
\begin{figure}
\centering
\includegraphics[width=1.0\textwidth]{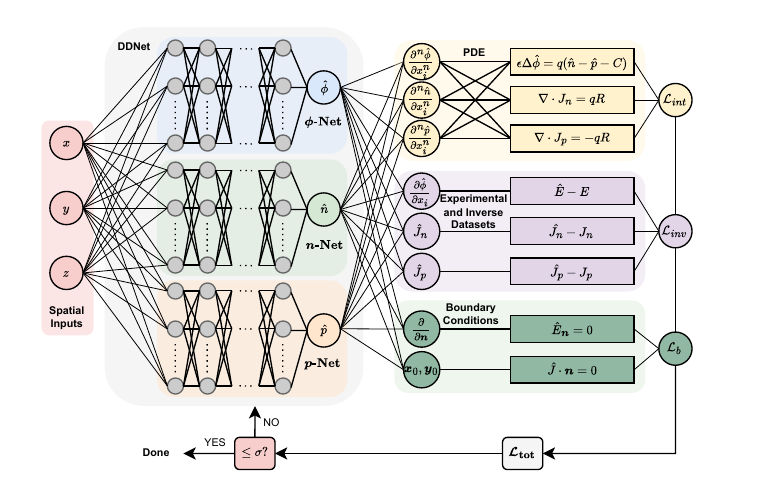}
\caption{The Drift-Diffusion Network (DDNet) architecture for forward semiconductor device modeling. The spatial inputs are passed to the three subnetworks $\phi$-Net, $n$-Net, and $p$-Net, which learn the three functions $\hat{\phi}$, $\hat{n}$, and $\hat{p}$, respectively. The three functions are then used to satisfy the physical constraints imposed by the Poisson-drift-diffusion equations via PDE and boundary conditions. If experimental data is available, it can be employed to further inform the DDNet solution. The PDE, boundary, and experimental residuals are then evaluated with a suitable loss function metric, which is then used to train DDNet using backpropagation. The process is repeated until a certain number of iterations (epochs) have passed or the loss has decreased below a threshold error $\sigma$.}\label{fig:fig1}
\end{figure}
DDNet solves the general stationary drift-diffusion system for $\phi,n,$ and $p$:
\begin{align}
\epsilon \Delta \phi &= q(n - p - C), \nonumber \\
\nabla \cdot \pmb{J}_n &= q R, \nonumber \\
\nabla \cdot \pmb{J}_p &= -q R, \nonumber \\
\pmb{J}_n &= q\mu_n(U_t \nabla n - n \nabla \phi), \nonumber \\
\pmb{J}_p &= -q\mu_p(U_t \nabla p + p \nabla \phi),
\label{eq:DD}
\end{align}
The Poisson-drift-diffusion equations describe charge transport in semiconductors by coupling the Poisson equation for the electrostatic potential $\phi$ to the continuity equations for the electrons ($n$) and holes ($p$) concentrations. To solve Eq.~\ref{eq:DD} we must specify the device permittivity $\epsilon$ and doping profile $C=N_d^+-N_a^-$. $C$ is the distribution of dopants within a semiconductor material, and it determines the electrical properties of the device. $\pmb{J}_n,\pmb{J}_p$ denote the electron and hole current densities, respectively. Their divergences are linked to the generation–recombination term $R$ that represents the net rate of electron–hole pair creation or annihilation per unit volume. In the simulations that follow, we include the Shockley-Read-Hall recombination term, but DDNet has the flexibility to incorporate other mechanisms such as Auger or radiative effects. Charge transport is dominated by two phenomena. Charges driven by the electric field give rise to the drift currents while gradients in charge density originate diffusion currents, with mobilities and diffusivities linked via the Einstein relation~\cite{1985Markowich, sze_physics_2021}. Solving the drift-diffusion model in Eq.~\ref{eq:DD} requires the specification of appropriate boundary conditions, which could be Dirichlet, Neumann, or mixed depending on specific device structures. DDNet is designed to simultaneously learn the scaled functions $\hat{\phi}$, $\hat{n}$, and $\hat{p}$~\cite{1985Markowich} (see the ``Methods" section).

\subsubsection{The DDNet architecture}\label{sec:subsec1}
As illustrated in Fig.~\ref{fig:fig1}, the DDNet architecture comprises three coupled fully-connected deep neural networks (FCNN), namely the $\phi$-Net, $n$-Net, and $p$-Net. In this work, we introduce a logarithmic FCNN architecture for the $n$-Net and $p$-Net subnetworks, which enables the accurate approximation of the large dynamic ranges of the solutions, often spanning twelve to eighteen orders of magnitude.
The networks are trained on randomly sampled collocation points in the simulation domain $\Omega$ and its boundary $\partial\Omega$, where the PDE residuals, boundary conditions, and (if applicable) inverse constraints are evaluated and minimized via a cumulative mean-squared error loss. 
The PDE and boundary conditions are combined in a cumulative loss function $\mathcal{L}_{tot}(\tilde{\theta}) = \mathcal{L}_{\text{int}}(\tilde{\theta}) + \mathcal{L}_{\text{BC}}(\tilde{\theta})$ where each term corresponds to a different interior or boundary PDE conditions:
\begin{align}
    \mathcal{L}_{\text{int}}(\tilde{\theta}) &= \mathcal{L}_{\text{Poisson}}(\tilde{\theta};\mathcal{N}_{\text{int}}) +  \mathcal{L}_{\pmb{J}_n}(\tilde{\theta};\mathcal{N}_{\text{int}}) + \mathcal{L}_{\pmb{J}_p}(\tilde{\theta};\mathcal{N}_{\text{int}}) \nonumber \\
    \mathcal{L}_{\text{BC}}(\tilde{\theta}) &= \mathcal{L}_{\hat{\phi},\text{BC}}(\tilde{\theta};\mathcal{N}_{\text{BC}}) +  \mathcal{L}_{\hat{n},\text{BC}}(\tilde{\theta};\mathcal{N}_{\text{BC}}) + \mathcal{L}_{\hat{p},\text{BC}}(\tilde{\theta};\mathcal{N}_{\text{BC}})
    \label{eq:loss_tot}
\end{align}
Each loss term $\mathcal{L}_{t}$ is computed using a mean squared error formula to minimize the residual error on $\mathcal{N}_{i}$ collocation or boundary points using the ADAM optimizer~\cite{kingma_adam_2017}:
\begin{equation}
    \mathcal{L}_{t} = \frac{1}{N_i} \sum_{n=1}^{N_i} \left|\left| \mathfrak{L}\left[ u_\theta(x_i^{(n)}, t_i^{(n)}) \right]\right|\right|^2_2
\end{equation}
Here, $\mathfrak{L}$ is a differential operator on a given DDNet output function $u_\theta(x_i^{(n)}, t_i^{(n)})$. In the inverse problem of Section~\ref{sec:subsec4}, we include an additional loss term to inform the training of the desired target performance:
\begin{equation}
    \mathcal{L}_{inv}(\tilde{\theta};\mathcal{N}_{inv}) =\frac{1}{|\mathcal{N}_{inv}|}\sum_{(x,y)\in\mathcal{N}_{inv}} \left|\left| \hat{E} - E_{\text{target}}\right|\right|^2_2
\end{equation}
After the optimization step is completed,  the network parameters $\tilde{\theta}$ are updated using backpropagation, and the process is repeated until a threshold error is achieved, or the training routine has reached a set number of epochs. For all the examples that follow, the loss functions of DDNet achieve values lower than $10^{-4}$, without the need for device- or geometry-specific fine-tuning, as shown in Table S2 and Figure S4 of the Supplementary Information.
\begin{figure}
\centering
\includegraphics[width=0.7\textwidth]{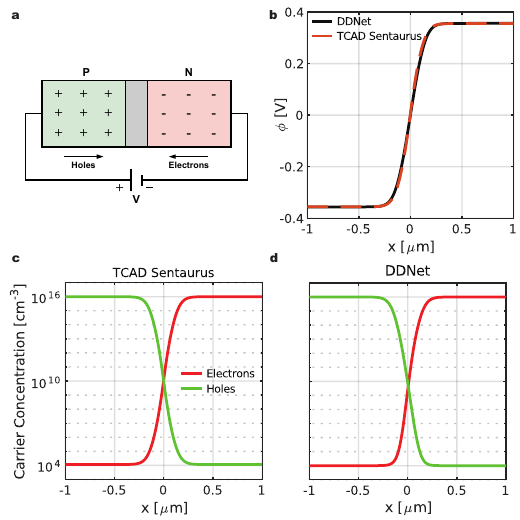}
\caption{One-dimensional forward device simulation with DDNet. a, Schematics of the $pn$ junction geometry employed for validating the network's forward modeling capabilities. b, Overlayed plot of TCAD Sentaurus and DDNet’s reconstruction of the electric potential. c, d, Comparison of the densities of electrons and holes simulated by TCAD Sentaurus (left) and DDNet (right). The relative $L_1$ errors are below 2\% for all cases.}\label{fig:fig2}
\end{figure}

We present here prototypical examples of semiconductor device configurations in order to demonstrate the capabilities of the developed DDNet framework and we compare its predictions with those obtained from state-of-the-art commercial solvers. The results were compared using the relative $L_1$ error defined in Section 2 of the Supplementary Information. The following results comprise both forward and inverse design problems, including simulations of 1D and 2D homogeneous silicon junctions under varying bias conditions. 
\subsection{One-dimensional \texorpdfstring{$p$-$n$}{p-n} junction}\label{sub:subsec2}
We first simulate a silicon 1D $p$-$n$ junction at equilibrium using an abrupt doping profile. Despite having an analytical solution~\cite{1985Markowich,sze_physics_2021}, this case is computationally challenging for traditional PINNs due to the presence of a discontinuity in $C$. As shown in Fig.~\ref{fig:fig2}a, the junction forms at the interface between $n$-type and $p$-type regions. At thermal equilibrium, the distribution of carriers creates a depletion region and establishes the built-in potential $\phi_{bi}$. Dirichlet Ohmic boundary conditions are applied for all fields to ensure uniqueness of the solution. Fig.~\ref{fig:fig2}b compares the DDNet prediction of $\phi$ with TCAD Sentaurus results, showing excellent agreement in logarithmic scale. The carrier density profiles are shown in Fig.~\ref{fig:fig2}c--d, achieving relative $L_1$ errors below 2\%. We also simulated the same junction under forward and reverse bias, and included the results in the Supplementary Information Figure S3. DDNet well-captures the evolution of potential and carrier profiles for various applied voltages. In particular, DDNet captures the expected shift of the depletion region. Additionally, we simulate and include in the Supplementary Information Figure S3 an asymmetric ($p^+$-$n$) junction simulation, where the depletion region shifts toward the less-doped side.

\begin{figure*}[ht]
\centering
\includegraphics[width=1.0\textwidth]{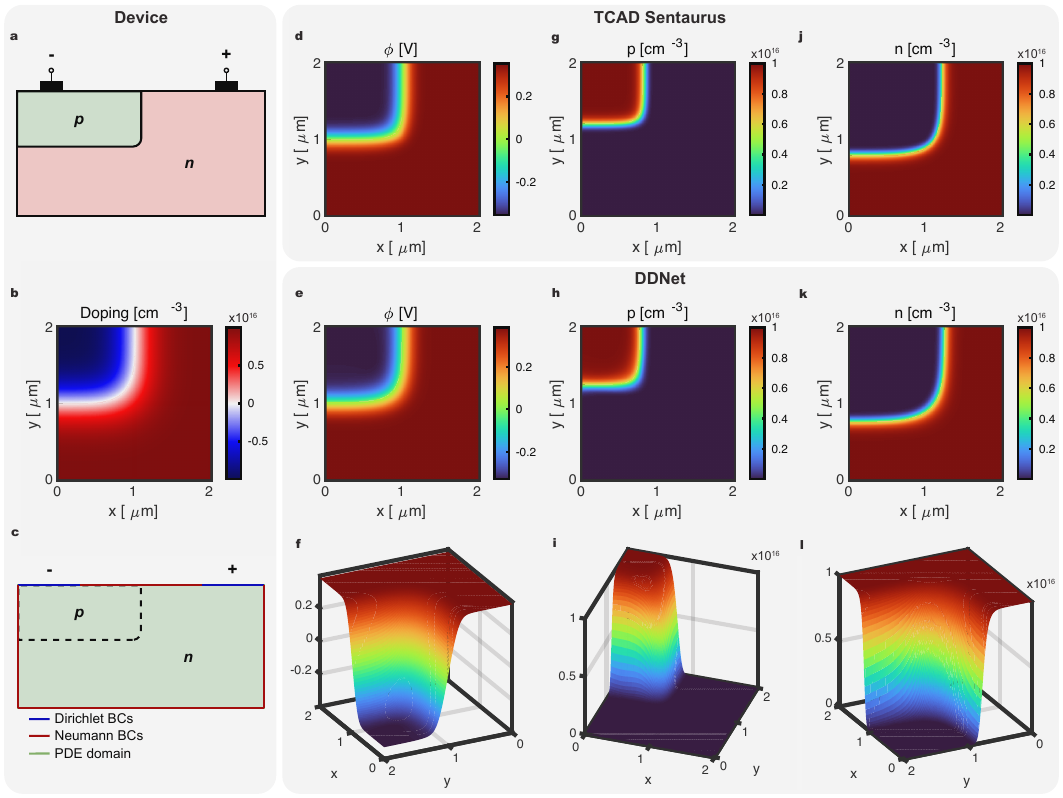}
\caption{Two-dimensional forward device simulation with DDNet. a, Schematics of the 2D $p$-$n$ junction geometry employed with its doping profile shown in b. c, The different boundary conditions used when training DDNet on this geometry. Each line segment corresponds to 3 additional PDE equations that DDNet solves simultaneously. d, g, l, The reference solution computed by using TCAD Sentaurus, with DDNet's solution shown below in panels e, h, and k. The 3D contour plots of the DDNet solutions are shown in panels f, i, and l. The integrated relative $L_1$ errors between d, g, j and e, h, k were below 5\%.}\label{fig:fig3}
\end{figure*}

\subsection{Two-dimensional \texorpdfstring{$p$-$n$}{p-n} junction}\label{sec:subsec3}
To demonstrate scalability to higher spatial dimensions, we simulate a 2D $p$-$n$ junction with the geometry shown in Fig.~\ref{fig:fig3}a and doping profile in Fig.~\ref{fig:fig3}b. The boundary conditions are schematized in panel c, where Dirichlet conditions are enforced on the device contacts and Neumann insulating conditions elsewhere, modeling zero current outflow. The DDNet predictions for $\phi$, $n$, and $p$ are shown in panels e, h, and k, respectively, and compared to TCAD Sentaurus results in panels d, g, and j. The 3D surface plots (panels f, i, l) confirm accurate reconstructions, with the average relative $L_1$ errors with TCAD Sentaurus's solutions falling below 5\%. This example demonstrates robust scalability to higher dimensions, highlighting the expressivity of DDNet. In fact, the architecture, collocation points, and loss functions were left unchanged for the solution of the 2D problem.


\begin{figure*}[t]
\centering
\includegraphics[width=1.0\textwidth]{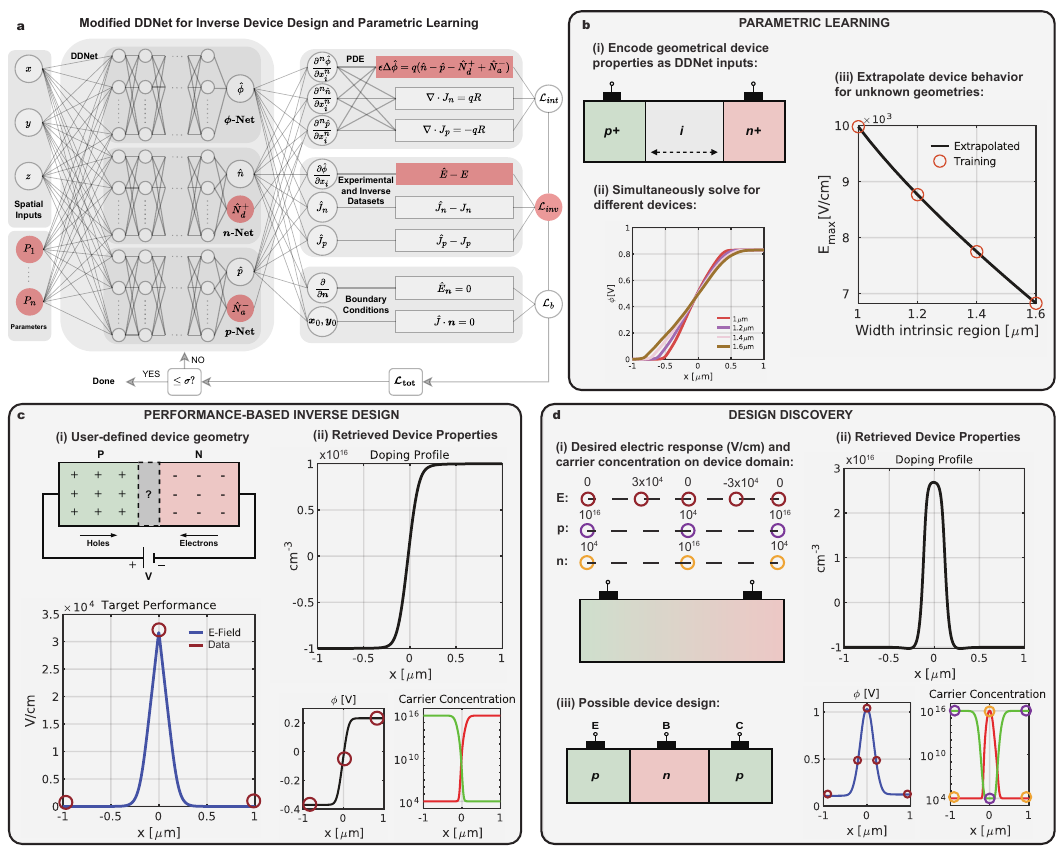}
\caption{Inverse design capabilities of DDNet. a, Modified DDNet architecture to tackle inverse design and parametric learning tasks. The input space has been expanded to accept device parameters $P_1,\dots,P_n$, and the output space of the $p$-Net and $n$-Net now enables the retrieval of the doping profile $\hat{C}=\hat{N}_d^++\hat{N}_a^-$. b, Example of parametric learning for a $pin$ junction, where DDNet extrapolates the maximal electric field behavior as a function of the intrinsically doped region width after training on 4 widths. c, Example of inverse design of a doping profile to achieve the maximal user-defined target performance from only three data points. d, Example of device discovery, where a $pnp$ BJT-like structure is retrieved to match the user's performance criteria on the electric field and carrier concentrations.}\label{fig:fig4}
\end{figure*}
\subsection{Inverse Design} \label{sec:subsec4}
\subsubsection{DDNet for inverse problems and parametric learning}
Oftentimes, in science and engineering, researchers are not only interested in the forward solution of a particular problem, but are keen to achieve the solutions of suitably defined inverse problems. That is, they are interested in the geometric properties or in the determination of fundamental parameters of the model that are consistent with a desired performance. This class of problems, known as \textit{inverse problems}, are often ill-posed (e.g., they do not have a unique solution) and are difficult and time-consuming to solve using traditional methods~\cite{burger_inverse_2004}. In particular, relevant inverse problems feature non-convex objective functions leading to a very complex solution landscape with multiple local minima, making it impossible to find global optimal solutions. For these reasons, they have traditionally been approached by constrained optimization methods that, through numerous iterations of a forward model, reconstruct simultaneously both the target function and the relevant physical properties of the system~\cite{chen_computational_2018, colton_inverse_2019}. Therefore, inverse problems are computationally more expensive than the corresponding forward problems due to the high dimensional nature of the parameter space of solutions that the solver needs to navigate in order to approach an optimal one. Despite their computational challenges, inverse problems have ubiquitous applications in science and engineering, ranging from MRI technologies to geophysics and remote sensing. However, in the context of semiconductor devices, a systematic approach to tackle relevant inverse problems is still missing, despite some recent attempts at extracting functional parameters using ML-based methods by Ling-Feng Mao et al. and Djordjevic et al.~\cite{djordjevic_inverse_2021, mao_addressing_2024}. Developing a systematic approach to solve inverse problems for semiconductor devices could enable a faster exploration of optimal device designs and properties. In addition, a realistic inverse design framework should be flexible enough to incorporate fabrication and material constraints effortlessly, thus enabling inverse design capabilities.
The 1D and 2D problems described thus far showed that DDNet can solve traditional forward semiconductor problems with precision comparable to state-of-the-art numerical solvers. However, DDNet is not intended to replace traditional, well-established forward numerical solvers, but rather to complement them by providing support for applications where they encounter difficulties. For this reason, we show in Fig.~\ref{fig:fig4} the range of inverse solutions that have been implemented and validated using DDNet so far. In Fig.~\ref{fig:fig4}a, we display the modified DDNet architecture developed to solve inverse problems. First, by directly encoding relevant device parameters within the input training vector of DDNet, here schematically represented by $P_1,\dots,P_n$ in panel a, we achieve parametric learning and robust extrapolation of device properties. For example, in the case of a $p$-$i$-$n$ diode geometry, used extensively as  photodetectors, we set up a training simulation where the width of the intrinsic region was incorporated as an additional learning parameter to DDNet, along with its spatial dimension. By simultaneously solving the drift-diffusion equations for four different doping conditions, we could then extrapolate the functional response of the electric field within the device region, which has been plotted against the width of the intrinsic region in panel b.
In addition, we show in panels c and d examples of performance-based inverse design discovery. In the first example, we aim at finding the slope of the doping profile $C(x)$ for a $p$-$n$ junction that features a desired electric field distribution specified at several user-defined points inside the device domain. This setup corresponds to a genuine inverse design problem, as opposed to simpler parameter retrieval problems~\cite{riganti_auxiliary_2023}. To solve inverse problems, we modify the $n$-Net and $p$-Net subnetworks to include the additional outputs $\hat{N}_d^+$ and $\hat{N}_a^-$, respectively, which specify the doping concentration of the inversely retrieved device structure $\hat{C}(\pmb{x})$.

Furthermore, we replace Dirichlet conditions on $\phi$ with a small set of target electric field data points, which are shown with red circles in panel a. An additional inverse loss term is added, allowing DDNet to learn both the doping profile and the forward solution simultaneously. Results for this first example are shown in panel c: the recovered doping profile has the appropriate slope to reproduce the desired target performance, while the predicted electric potential and carrier densities remain accurate. This example shows that, despite the ill-posed nature of this inverse problem, DDNet recovers a realistic doping profile from the family of possible solutions. 

As a second example, we attempt to inversely design an unknown device configuration (i.e., doping profile) following the same principle, as shown in panel d. This time we select $N=5$ points that we would like to match to a different electric field distribution on the device domain. In addition, we select 3 more points that we want to match with a specified carrier concentration of $n$ and $p$. However, we might be unaware of the existence of a device with such features. Using the same DDNet architecture, we recover a doping profile that exhibits the desired electric field distribution, as shown by the results in panel d. A posteriori, we would learn that this structure represents a BJT-like (or $p$-$n$-$p$ junction) device with symmetric doping. However, the solution of the inverse design problem was guided solely by the target characteristics, not the material geometry, under the constraint of the total length. Rather, the material properties, i.e., the doping profile, were derived during the training of DDNet to satisfy the drift-diffusion equations. For both examples c and d, we have overlaid the position of the target data points (red circles) on the reconstructed forward simulation of the electric potential.
Furthermore, we demonstrate the capabilities of DDNet in relation to two analytically intractable problems of inverse semiconductor device design. 
\begin{figure}
\centering
\includegraphics{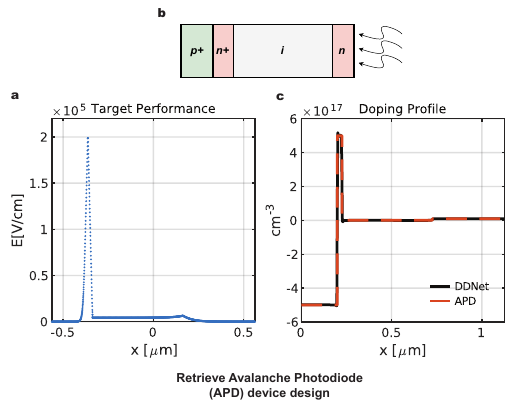}
\caption{Inverse design of APDs and binarized structures using DDNet. a, Multiscale electric field target performance dataset for a prototypical RF sensing device. b, DDNet's inverted doping profile with a comparison to a true APD doping and corresponding APD device schematic. }\label{fig:figAPD}
\end{figure}
\subsubsection{Inverse design of an avalanche photodiode (APD)}
We first attempt the retrieval of the doping profile for a $1\mu \text{m}$ avalanche photodiode (APD), characterized by a high electric field response at equilibrium, which under reverse bias enables the amplification of small photogenerated signals. The set-up of this inverse design problem is presented in  Fig.~\ref{fig:figAPD}a, where the target electric field response is plotted over the device domain. After 3 minutes of training time, DDNet performs a full inversion of the doping profile of an APD, as shown in panel b. Interestingly, DDNet retrieves the $25$ nm thick $n^+$ doped region needed to enhance the electric field response after the avalanche region between the intrinsic ($i$) and $p^+$ doped regions, as shown schematically in b. This doping inversion was performed only with prior knowledge of the desired electric response of the device subject to the drift-diffusion equations, and DDNet's training required no additional time or computational costs compared to the corresponding 1D forward simulations.
\begin{figure*}
\centering
\includegraphics[width=1\textwidth]{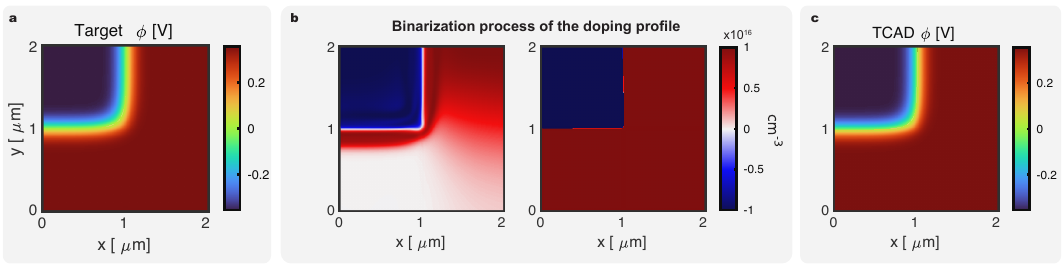}
\caption{Inverse design of a binarized structure using DDNet. a, Target electric potential inversion dataset. b, Binarization retrieval process during DDNet's inverse design, where an impractical grated doping profile (left) is slowly binarized during the process until a binarized one is obtained (right). c, Forward simulation with the binary doping profile shown in d using TCAD Sentaurus, in close agreement with the target.}\label{fig:fig2Dinv}
\end{figure*}
\subsubsection{Inverse design of binary doping with fabrication constraints}
As our last example, we demonstrate how to include fabrication constraints during the inverse design process. In Fig.~\ref{fig:fig2Dinv}a we show the two-dimensional plot of the electric potential from the junction discussed in \ref{sec:subsec3}. This electric potential profile was obtained using the continuous doping profile shown in Fig.~\ref{fig:fig3}b. However, we now want to inversely obtain a binarized doping profile that produces a device performance as close as possible to the original (non-binarized) situation. To achieve this, we set up our full inverse design of the doping profile by providing DDNet the 2D data of the target electric potential, and only specifying the values of $\hat{p}$ and $\hat{n}$ at the source and drain contacts. A first solution, which is obtained with no additional constraints imposed on $\hat{C}$, is shown in the first figure of panel d. This figure shows a graded doping profile that, despite satisfying the Poisson equation, has limited applicability, since it cannot be practically implemented. To avoid retrieving impractical solutions, we borrow an inverse design idea from topological optimization~\cite{christiansen_inverse_2021}, which has recently been employed in the inverse design of photonic structures~\cite{riganti_multiscale_2025}. Namely, we slowly binarize the DDNet output $\hat{C}$ during the training process with a custom thresholding function $\sigma_{cst}(x,t)=\frac{\gamma}{1+e^{-\xi(t)\cdot x}}+\eta$ to ensure that the output does not deviate from our fabrication requirements. Here, $\sigma_{cst}(x,t)$ is explicitly time-dependent since its slope $\xi(t)$ slowly increases during the training process, allowing for a smooth solution in the early epochs. As the training proceeds, the slope becomes sharper until the doping converges to a perfectly binary solution. The result is shown on the right of panel d, where a fully binary doping profile has been obtained. To validate the accuracy of the newly retrieved structure, we perform a forward simulation in TCAD Sentaurus and plot $\phi$ in Fig.~\ref{fig:fig2Dinv}c. The relative $L_1$ error between the target and TCAD $\phi$ was lower than 1\%. Similar to the 1D case, the 2D inversion problem added no computational overhead compared to the forward problem.
These inverse design examples highlight one of the key strengths of DDNet: the ability to solve inverse PDE problems without incurring additional computational cost or requiring large datasets. 
\section{Discussion}
We introduced DDNet, a unified physics-informed deep learning framework to solve forward, parametric, and inverse design problems in semiconductor engineering. As an inverse design tool, DDNet usually trains on synthetic data, but has the infrastructure to augment its predictive power by incorporating experimental or target performance datasets at minimal computational overhead. Furthermore, the architecture shown here can be fine-tuned to solve semiconductor problems for different materials and geometries, including heterostructures.
The novel framework presented complements and expands the capabilities of traditional forward solvers and data-driven machine learning algorithms, demonstrating inverse design capabilities and functionalities not explored before. 
Future research will focus on scaling and generalizing DDNet to more complex device geometries, multiple materials, and coupled multi-physics applications, such as optical and thermal transport. 

\section{Methods}

\subsection{Boundary Conditions}
\subsubsection{Ohmic Contacts}
At ohmic contacts, the device is assumed to be in thermal equilibrium with the external circuit, and carrier concentrations are pinned to their equilibrium values specified by the local doping concentration. Considering the case of non-degenerate statistics for simplicity, the ohmic contacts as boundary conditions at $x = x_c$ are given by:
\begin{align}
\phi(x_c) &= \phi_0-V_a, \\
n(x_c) &= n_i \exp\left( \frac{\phi(x_c)}{U_t} \right), \label{eq:bc_n_ohmic} \\
p(x_c) &= n_i \exp\left( -\frac{\phi(x_c)}{U_t} \right), \label{eq:bc_p_ohmic}
\end{align}
\noindent where $\phi_0 = U_t \ln\left( \frac{n}{n_i} \right)$ for \textit{n}-type contact or $\phi_0 = -U_t \ln\left( \frac{p}{n_i} \right)$ for \textit{p}-type contact. $V_{\text{a}}$ is the externally applied potential at the contact, $n_i$ is the intrinsic carrier concentration, and $U_t = kT/q$ is the thermal voltage. These expressions follow from the condition of charge neutrality $n - p = C$, and the Boltzmann approximation $n = n_i \exp(\phi/U_t)$, $p = n_i \exp(-\phi/U_t)$.
\subsubsection{Insulating Boundaries}
On insulating boundaries (e.g., semiconductor–oxide or symmetry planes), no current is allowed to flow across the surface, and the normal component of the electric displacement must vanish. These Neumann-type conditions are:
\begin{align}
\pmb{D} \cdot \pmb{n} &= -\epsilon \nabla \phi \cdot \pmb{n} = 0, \label{eq:bc_D_neumann} \\
\pmb{J}_n \cdot \pmb{n} &= 0, \label{eq:bc_Jn_neumann} \\
\pmb{J}_p \cdot \pmb{n} &= 0, \label{eq:bc_Jp_neumann}
\end{align}
where $\pmb{n}$ is the outward unit normal to the boundary. 
\subsection{Equations rescaling and logarithmic treatment}
The modified drift-diffusion equations used to train DDNet are obtained by rescaling Eq.~\ref{eq:DD} according to the scaling factors in \cite[Table 2.4.1]{1985Markowich}:
\begin{align}
\lambda^2 \Delta \hat{\phi} &= \hat{n}-\hat{p}-C \nonumber \\
\nabla \cdot \pmb{J}_n  &= R \nonumber \\
\nabla \cdot \pmb{J}_p  &= -R \nonumber \\
\pmb{J}_n &= \mu_n(\nabla \hat{n} - \hat{n} \nabla \hat{\phi} ) \nonumber \\
\pmb{J}_p &= -\mu_p(\nabla \hat{p} + \hat{p} \nabla \hat{\phi} ) 
\label{eq:DD_scaled}
\end{align}
In Eq.~\ref{eq:DD_scaled}, we have introduced the scaling constant $\lambda=\lambda_D/\ell$, where $\lambda_D=\sqrt{\frac{\epsilon_s U_t}{q \tilde{C}(x)}}$ is the Debye length of the device, and $\ell$ is a characteristic device length chosen such that it is of the same order of magnitude as the diameter of the semiconductor simulation domain $\Omega$, $\ell = O(\text{diameter}(\Omega))$. A complete description of the scalings and variables can be found in Section 1 and Table S1 of the Supplementary Information.
The equations in~\ref{eq:DD_scaled} are obtained to aid with the numerical treatment of the functions $\hat{\phi}$, $\hat{n}$, and $\hat{p}$~\cite{1985Markowich}. In particular, this formulation helps deal with the large dynamic ranges of $\hat{n}$ and $\hat{p}$, which typically span between twelve and sixteen orders of magnitude for silicon, but they can reach up to forty orders of magnitude for materials like GaN. To do so, $\hat{n}$ and $\hat{p}$ are rescaled by $\tilde{C}$, defined as $\tilde{C}=O(C_{\text{max}})$, which ensures that the maximum value of $\hat{n}$ and $\hat{p}$ is unity, while $\hat{\phi}$ is rescaled by $U_t$. 

To capture the full dynamic ranges of $\hat{n}$ and $\hat{p}$ we rescale the outputs according to $\hat{p}/\tilde{C}$ and $\hat{n}/\tilde{C}$ and train according to the newly defined variables -log($\hat{p}/\tilde{C}$) or -log($\hat{n}/\tilde{C}$), which compress their dynamic range. Finally, the outputs are exponentially transformed through a hard constraint~\cite{lu_physics-informed_2021} as $e^{-\text{log}(\hat{n}/\tilde{C})}$ and $e^{-\text{log}(\hat{p}/\tilde{C})}$, thereby recovering the full dynamic range of the solutions. We have compared the training using the logarithmic scaling with a traditional PINN and included the analysis in Section 3 of the Supplementary Information. We show that the traditional PINN framework achieves precision comparable to DDNet for the electrostatic potential, as shown in Figure S1. However, Figure S2 shows that the traditional PINN fails to capture the full dynamic range for $n$ and $p$.

\subsection{Training details}
To train DDNet, we typically use $2^{12}=4096$ uniform random collocation points on the PDE domain and $2^{8}=256$ uniform random collocation points on the boundary for a total of $N=40,000$ epochs. Additionally, we introduce a resampling routine during training that re-generates the collocation points every $N/f$ epochs, where $f$ is an adjustable frequency set to 100 by default. We have found this technique to improve with convergence and precision without introducing a significant additional overhead. All the DDNet codes and training routines have been developed in-house using the TensorFlow~\cite{tensorflow2015-whitepaper} machine learning package, and the scripts ran on a single L40S GPU of the Boston University Shared Computing Cluster (SCC). 

\section*{Supplementary information}
We have included additional data on DDNet training and examples, network architecture comparisons, convergence times, and loss function convergence in the Supplementary Information.

\section*{Funding} This work was supported by the U.S. Army Research Office, RF-Center managed by Dr. T. Oder (Grant \#W911NF-22-2-0158).

\section*{Author Contributions} E.B. and L.D.N. obtained the grant and initiated this project. L.D.N. conceived the idea and oversaw the technical research.  R.R. designed the model, trained the model, and obtained the results. R.R. and L.D.N. wrote the main manuscript with valuable inputs from all the authors.  E. B. and M.G.C.A. helped develop relevant application examples. All authors reviewed the manuscript.

\section*{Competing Interests} The authors declare no competing interests.


\bibliography{DD}

\end{document}


\title[Article Title]{Supplementary Information - DDNet: A Unified Physics-Informed Deep Learning Framework for Semiconductor Device Design}


\author[]{\fnm{Roberto} \sur{Riganti}}

\author[]{\fnm{Matteo G. C.} \sur{Alasio}}

\author[]{\fnm{Enrico} \sur{Bellotti}}

\author[]{\fnm{Luca} \sur{Dal Negro}}





\maketitle
\newpage

\tableofcontents

\newpage

\section{Drift-diffusion equations scalings}\label{SI:sec2}
We rewrite here the drift-diffusion equations introduced in the main text:
\begin{align}
\epsilon \Delta \phi = q(n-p-C) \nonumber \\
\nabla \cdot \pmb{J}_n  = q R \nonumber \\
\nabla \cdot \pmb{J}_p  = -q R \nonumber \\
\pmb{J}_n = q\mu_n(U_t \nabla n - n \nabla \phi) \nonumber \\
\pmb{J}_p = -q\mu_p(U_t \nabla p + p \nabla \phi) 
\label{eq:DD}
\end{align}
where:
\begin{itemize}
    \item $\epsilon$ is the material permittivity, which for Silicon is $\epsilon=11.7*8.85*10^{-14} \frac{C^2 s^2}{kg\cdot cm^3}$
    \item $q=1.6*10^{-19}C$ is the electron charge
    \item $R_{SRH} = \frac{np-n_i^2}{\tau_p^l(n+n_i)+\tau_n^l(p+n_i)}$ is the generation recombination rate, which represents the net rate at which electron-hole pairs are created or annihilated per unit volume per unit time. For this study, we considered the Shockley-Read-Hall (SRH) process, which models two-particle-transitions. However, DDNet has the flexibility to incorporate other mechanisms such as Auger or radiative effects. $\tau_n^l$ and $\tau_p^l$ are the electron and hole lifetimes, typically of the order of microseconds for Silicon. 
    \item $n_i$ is the intrinsic carrier concentration, that is the number of free electrons (and holes) per unit volume in a pure (undoped) semiconductor at thermal equilibrium. For Si, we used $n_i \approx 10^{10}\text{cm}^{-3}$.
    \item $\mu_n=1417\;cm^2/V/s,\;\mu_p=470.5 \;cm^2/V/s$ are the electron and hole constant low-field mobilities. They describe how quickly electrons and holes move in response to an electric field inside a semiconductor.
    \item $U_t = \frac{k_B T}{q} \approx 0.0259$V is the thermal voltage for Silicon at room temperature, where $T=300K$ and $k_B\approx1.38*10^{-23}$ is Boltzmann's constant.
\end{itemize}

To introduce the scaled equations, we report in Table~\ref{tabSI:scales} the scaling factors adapted from Table 2.4.1 of Ref.~\cite{2012Markowich}. In this table, we have written TBC for scaling factors that are problem dependent, or "to be computed."
\begin{table}[h]
\caption{Scaling constants}\label{tabSI:scales}%
\begin{tabular}{@{}llll@{}}
\toprule
Variable & Meaning & Scaling factor & Value\\
\midrule
x,y,z    & spatial variables & $\ell$ & $O(\text{diam}(\Omega))$  \\
$\phi$    & electronic potential & $U_t$ & TBC \\
$n$    & electron concentration & $\tilde{C}$ & $O(\text{max}[C])$  \\
$p$    & hole concentration & $\tilde{C}$ & $O(\text{max}[C])$ \\
$C$    & doping profile  & $\tilde{C}$ & $O(\text{max}[C])$ \\
$n_i$    & intrinsic carrier concentration & $\tilde{C}$ & $O(\text{max}[C])$ \\
$\mu_n,\; \mu_p$    & mobilities  & $\tilde{\mu}$ & max($\mu_p,\mu_n$) \\
$\tau_n^l,\;\tau_p^l$    & lifetimes  & $\frac{\ell^2}{\tilde{\mu}U_t}$ & TBC \\
\botrule
\end{tabular}
\end{table}

The scaled drift-diffusion equations become:
\begin{align}
\lambda^2 \Delta \hat{\phi} = \hat{n}-\hat{p}-C \nonumber \\
\nabla \cdot \pmb{J}_n = R \nonumber \\
\nabla \cdot \pmb{J}_p  = -R \nonumber \\
\pmb{J}_n = \mu_n(\nabla \hat{n} - \hat{n} \nabla \hat{\phi} ) \nonumber \\
\pmb{J}_p = -\mu_p(\nabla \hat{p} + \hat{p} \nabla \hat{\phi} ) 
\label{eq:DD_scaled}
\end{align}
where every quantity has been scaled by the values in Table~\ref{tab1}. $\lambda=\lambda_D/\ell$, where $\lambda_D=\sqrt{\frac{\epsilon_s U_t}{q \tilde{C}}}$ is the Debye length.

\section{The relative $L_1$ error norm}\label{SI:L1}
In the article, we employ the relative $L_1$ error to quantify the accuracy of DDNet's predictions against TCAD Sentaurus. The relative $L_1$ error is defined as:
\begin{equation}
    L_1 = \frac{\sum_i|\hat{y}-y_{true}|}{\sum_i|y_{true}|}
\end{equation}
Here, $\hat{y}$ corresponds to DDNet's prediction and $y_{true}$ corresponds to TCAD Sentaurus's solution.

\newpage

\section{DDNet architecture - comparison between traditional and logarithmic formulation}\label{SI:architecture}
Traditional PINN architectures are often built using relatively simple feed-forward fully-connected neural networks (FCNN)~\cite{chen_physics-informed_2020, lu_deepxde_2021, raissi_physics-informed_2019}. However, in recent years, more exotic architectures have been introduced in order to tackle the limitations of these traditional architectures in solving multiscale PDE problems~\cite{riganti_multiscale_2025}, boundary conditions in inverse design~\cite{lu_physics-informed_2021}, or integro-differential equations~\cite{riganti_auxiliary_2023}, to name a few. Similarly, when solving the drift-diffusion equations, the immediate challenge posed to traditional architectures is the wide dynamic range of the solutions of the carrier concentrations, often exceeding twelve orders of magnitude in even the most simple examples. In this section we will demonstrate the need to employ a logarithmic PINN architecture in order to capture the dynamic ranges of $n$ and $p$ fully. For completeness, we will perform our analysis as follows: we will use two PINN architectures with the same depth (4 layers) and width (64 neurons per layer). However, one architecture will directly output the functions $n$ and $p$ and train on the PDE and boundary conditions described in the main text. We will call this first variation of the DDNet architecture the \textit{traditional architecture}. The second one, that is, the DDNet architecture employed in the manuscript, will output $-\text{log}(\hat{p}/\tilde{C})$ and $-\text{log}(\hat{n}/\tilde{C})$ and will transform both outputs into $e^{-\text{log}(\hat{n}/\tilde{C})}$ and $e^{-\text{log}(\hat{p}/\tilde{C})}$ through a hard constraint~\cite{lu_physics-informed_2021}. The negative sign is due to the scaling by $\tilde{C}=10^{16}$ (for this example), which shifts the dynamic range from $10^4-10^{16}$ to $10^{-12}-10^0$. We will call this DDNet variation the \textit{logarithmic architecture}. The scaled equations and variables presented in Section~\ref{SI:sec2} of the Supplementary Information were employed for both trainings, with identical loss functions, collocation points, and solver employed.
\begin{figure}[h]
\centering
\includegraphics[width=0.7\textwidth]{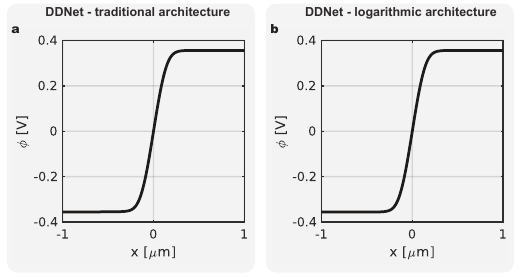}
\caption{a) Comparison between expressivity of the electrostatic potential $\phi$ solutions using a traditional PINN architecture and the original logarithmic formulation implemented in this work. The solutions are identical.}\label{fig:phiSIcomp}
\end{figure}
\newpage
As expected, after training for 40,000 epochs ($\sim3$ minutes), both architectures converge to a solution that captures the electrostatic potential $\phi$, as shown in Figure~\ref{fig:phiSIcomp}. This is expected because the electrostatic potential function of a $p$-$n$ junction has a relatively low dynamic range ($=\phi_{bi}$ at thermal equilibrium), and a traditional FCNN architecture should not have any issues in representing the function. However, for the carrier concentrations, we see that the traditional architecture breaks down. 

\begin{figure}[h]
\centering
\includegraphics[width=0.65\textwidth]{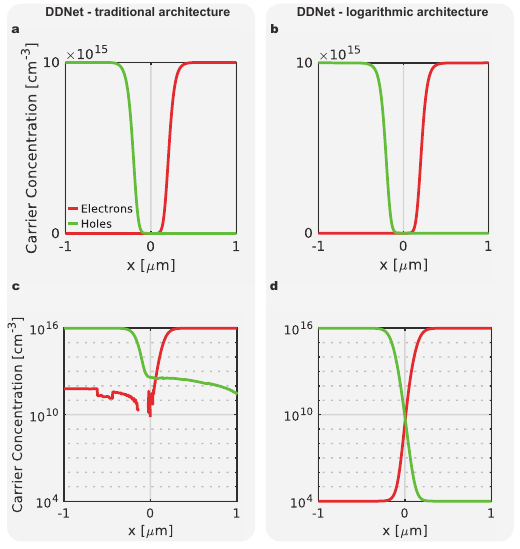}
\caption{Expressivity comparison between the carrier density solutions using a traditional PINN architecture and the original logarithmic formulation implemented in this work. The solutions are seemingly identical when the carrier concentrations are not plotted on a logarithmic scale, as shown in panels a and b. However, on a logarithmic scale, the traditional PINN architecture misses most of the dynamic range of the solutions, as shown in panel c, while the logarithmic architecture can capture the full dynamic range, as shown in panel d.}\label{fig:pnSIcomp}
\end{figure}

Panels a and b of Figure~\ref{fig:pnSIcomp} show apparent agreement between the carrier concentration solutions obtained using the traditional and logarithmic architectures. Both plots are shown in a linear y-scale, which is misleading when dealing with functions with large dynamic ranges. In fact, when the plots are shown in a logarithmic y-scale in panels c and d, we can immediately see that the traditional architecture can only appropriately capture 3-4 orders of magnitude in the solution. On the other hand, the logarithmic formulation captures the full dynamic range. In conclusion, despite traditional architectures enable an accurate prediction of the largest orders of magnitude of the carrier concentration solutions, only a logarithmic formulation successfully spans the entire dynamic range.

\newpage

\section{Additional examples}\label{SI:add_examples}

We report below additional validation examples of DDNet for the 1D $p$-$n$ junction presented in the main text. The solutions shown here were obtained with the same DDNet architecture, but they were trained with different boundary conditions or doping profiles. For panels a-f, we display the widening of the depletion region caused by an increased reverse bias voltage, and for panels g-h we show the shift of the depletion region toward the less-doped side in a $p^+$-$n$ junction, which is characterized by an asymmetric doping profile of $N_d^-=10^{17},N_a^+=10^{16}$.

\begin{figure}[h]
\centering
\includegraphics[width=1\textwidth]{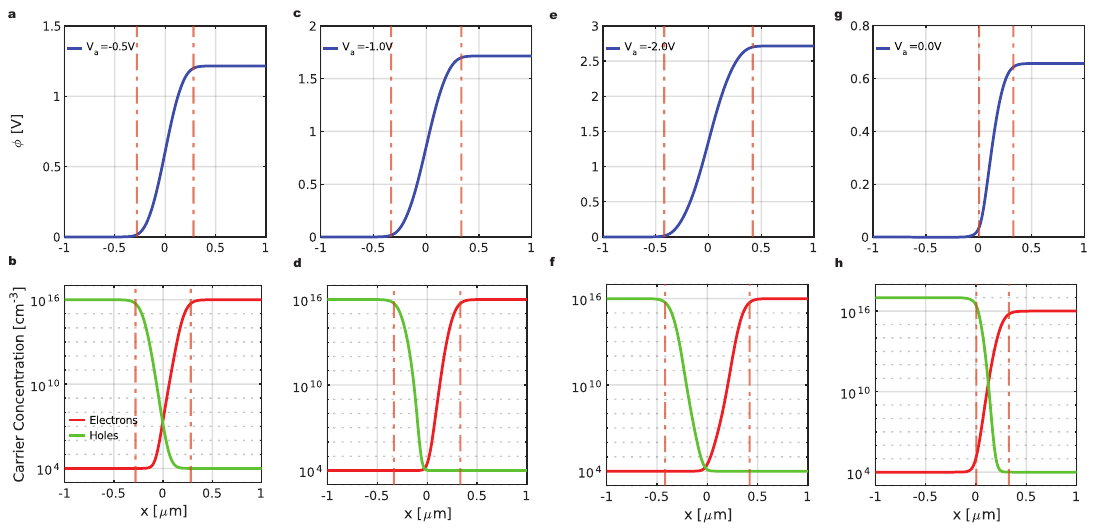}
\caption{$pn$ junctions simulations at different applied voltages and asymmetric doping profiles. a, c, and e show the reconstructed electric potential under different applied voltages, with the respective electron and hole densities shown in the line below, in panels b, d, and f, respectively. The vertical columns represent the predicted onset of the depletion region for the analytical step-function solution, which despite being different, it does not deviate too far from our error function doping profile. Finally, panels g and h show that DDNet also accurately captures the electric potential and electron and hole densities for asymmetrically doped $p$-$n$ junctions.
}\label{fig:fig3}
\end{figure}

\newpage

\section{Loss function convergence during training}\label{SI:loss}
All the DDNet trainings were performed on a single L40S GPU using the ADAM optimizer and a piecewise-constant learning rate decay starting from 0.01 and decreasing by a factor of 10 for a total of 8 times. 

We performed tests using both float32 and float64 data types in TensorFlow and did not notice any significant improvement on the resulting comparison with the TCAD Sentaurus solution, which was assumed to be the ground truth.

In the table below, we list the cumulative loss function values at the start and end of training for each study treated in the main text. We refer to each study by the figure number and lettered panel in which they appear, for simplicity. 
\begin{table}[h]
\caption{Cumulative training loss functions}\label{tab1}%
\begin{tabular}{@{}llll@{}}
\toprule
Figure & Loss function at the start of training & Loss function at the end of training \\
\midrule
2(b-d)  & 1e4 & 1e-6   \\
3a   & 1e4 & 1e-6  \\
3b   & 1e4 & 1e-6  \\
3c   & 1e4 & 1e-6  \\
3d   & 1e4 & 1e-6  \\
4    & 1e5 & 1e-4  \\
5    & 1e4 & 1e-6  \\
\botrule
\end{tabular}
\end{table}

\begin{figure}[h]
\centering
\includegraphics[width=1.0\textwidth]{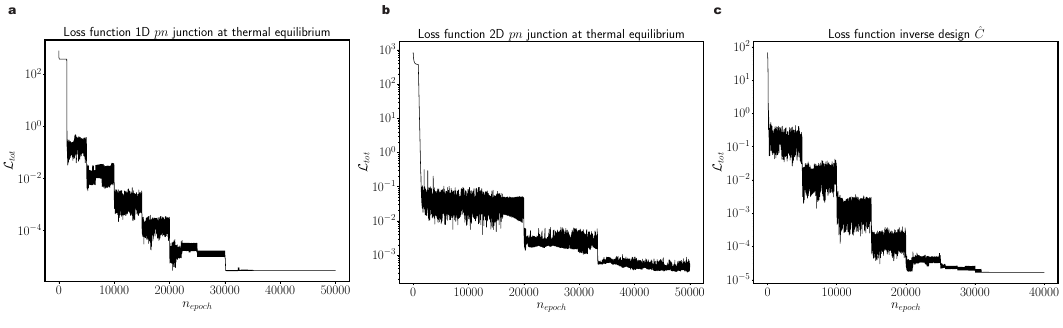}
\caption{Representative examples of the training loss function convergence for problems presented in the main text.}\label{fig:loss}
\end{figure}

In Figure~\ref{fig:loss} we show some representative examples of loss function convergence traces for problems presented in the main text. The loss function consistently decreases to values below $10^{-4}$, and we can distinctly see the impact of the stepwise constant decay.

\newpage

\section{GPU computation time}\label{SI:GPUs}
The DDNet implementation employed to produce the one-dimensional forward and inverse studies in the manuscript was tested to analyze its performance when running on different GPUs. The time reported here is the necessary time to reach convergence of the training loss, or 20,000 epochs. The fast convergence time for GPUs with capabilities above $6.0$ underscores the broad accessibility of this methodology for semiconductor device modeling and discovery~\cite{noauthor_nvidia_nodate}.
\begin{table}[h]
\caption{GPU time for a prototypical one-dimensional simulation with the DDNet framework}\label{tabSI:GPUtime}%
\begin{tabular}{@{}llll@{}}
\toprule
GPU Capability & GPU Type & Time forward or inverse simulation (min)\\
\midrule
8.9    & L40S & 2.5\\
8.0    & A100 80GB & 3.2\\
7.0    & V100-SXM2-16GB & 3.8\\
6.0    & P100-PCIE-16GB & 5.6\\
3.5    & Tesla K40m & 12\\
\botrule
\end{tabular}
\end{table}

\newpage




\newpage


\bibliography{DD}